\pdfoutput=1
\documentclass[review]{elsarticle}

\usepackage{xcolor}
\usepackage{hyperref}

\newcommand{\B}[1]{{\bm #1}}

\newcommand{\ds}{\displaystyle}

\newcommand{\ct}{\tau}

\usepackage{bm}
\usepackage{amsmath}
\usepackage{subfigure}

\newcommand{\vb}[1]{\mathbf{#1}}

\newcommand{\C}{\overline{C}_{nm}}
\newcommand{\N}{\mathcal{N}_{nm}}
\renewcommand{\P}{\overline{P}_{nm}}
\renewcommand{\S}{\overline{S}_{nm}}

\newcommand{\fracpar}[2]{\frac{\partial #1}{\partial #2}}

\usepackage{empheq}
\usepackage{gensymb}

\renewcommand{\a}{\alpha}
\renewcommand{\b}{\beta}
\renewcommand{\r}{\rho}
\renewcommand{\t}{\theta}

\usepackage{graphicx}
\usepackage{txfonts}

\journal{Acta Astronautica}

\begin{document}

\begin{frontmatter}

\title{Theory of Functional Connections and Nelder-Mead optimization methods applied in satellite characterization}

\author[it]{Allan K. de Almeida Jr \corref{cauthor}}
\ead{allan.junior@ua.pt}
\cortext[cauthor]{Corresponding author}

\author[inpe]{Safwan Aljbaae}
\ead{safwan.aljbaae@gmail.com}

\author[it,ucoimbra]{Timothée Vaillant}
\ead{vaillant@av.it.pt}

\author[unifesp]{Jhonathan M. Piñeros}
\ead{jhonathan.pineros@unifesp.br}

\author[it]{Bruno Coelho}
\ead{brunodfcoelho@av.it.pt}

\author[it]{Domingos Barbosa}
\ead{dbarbosa@av.it.pt}

\author[it]{Miguel Bergano}
\ead{jbergano@av.it.pt}

\author[superior]{João Pandeirada}
\ead{joao.pandeirada@av.it.pt}

\author[inpe]{Francisco C. Carvalho}
\ead{francisco.chagas@inpe.br}

\author[pernambuco]{Leonardo B. T. Santos}
\ead{leonardobarbosat@hotmail.com}

\author[inpe]{Antonio F. B. A. Prado}
\ead{antonio.prado@inpe.br}

\author[ubi]{Anna Guerman}
\ead{anna@ubi.pt}

\author[ucoimbra,imcce]{Alexandre C. M. Correia}
\ead{acor@uc.pt}

\address[it]{Instituto de Telecomunica\c{c}\~oes, Universidade de Aveiro, 3810-193 Aveiro, Portugal}
\address[inpe]{Postgraduate Division - National Institute for Space Research (INPE), São Paulo, Brazil}
\address[ucoimbra]{CFisUC, Departamento de Física, Universidade de Coimbra, 3004-516 Coimbra, Portugal}
\address[unifesp]{ICT-UNIFESP, Institute of Science and Technology, São José dos Campos, Brazil}
\address[superior]{Instituto Superior Técnico, Avenida Rovisco Pais 1, 1049-001 Lisboa Portugal}
\address[pernambuco]{Polytechnic School, University of Pernambuco, 50720-001 Recife, PE, Brazil}
\address[ubi]{University of Beira Interior, Covilhã, Portugal}
\address[imcce]{IMCCE, UMR8028 CNRS, Observatoire de Paris, PSL Université, 77 Avenue Denfert-Rochereau, 75014 Paris, France}

\begin{abstract}
The growing population of man-made objects with the build up of
mega-constellations not only increases the potential danger
to all space vehicles and in-space infrastructures (including space
observatories), but above all poses a serious threat to astronomy and
dark skies. Monitoring of this population requires precise satellite
characterization, which is is a challenging task that involves analyzing observational data such as position, velocity, and light curves using optimization methods. In this study, we propose and analyze the application of two optimization procedures to determine the parameters associated with the dynamics of a satellite: one based on the Theory of Functional Connections (TFC) and another one based on the Nelder-Mead heuristic optimization algorithm. 
    The TFC performs linear functional interpolation to embed the constraints of the problem into a functional. In this paper, we propose to use this functional to analytically embed the observational data of a satellite into its equations of dynamics. After that, any solution will always satisfy the observational data. 
  The second procedure proposed in this research takes advantage of the Nealder-Mead algorithm, that does not require the gradient of the objective function, as alternative solution. The accuracy, efficiency, and dependency on the initial guess of each method is investigated, analyzed, and compared for several dynamical models.
  These methods can be used to obtain the physical parameters of a satellite from available
observational data and for space debris characterization contributing to
follow-up monitoring activities in space and astronomical observatories.
\end{abstract}

\begin{keyword}
\texttt{astrodynamics, numerical methods, satellite characterization}
\end{keyword}

\end{frontmatter}


\section{Introduction}

The number of space activities, including the launch of constellations of satellites and the associated space debris population, has been steadily increasing, raising concerns over the impact on the sky quality for astronomical purposes.  This growing population of man-made objects not only increases the potential danger to all space vehicles and in-space infrastructures, from expensive astronomical space observatories to communications satellites, Earth Observation satellite constellations, Space Stations, 
and other spacecraft with humans aboard, but also severely impacts the quality of Earth's night sky and radio sky. This threat to the orbital space environment affects those using space assets and those who look through it as well as work within it \citep{2022NatAs...6..428L,2022AcAau.200..612B}.
In fact, since 2009, the International Astronomical Union (IAU)'s Resolution B5, ``In Defence of the Night Sky and the Right to Starlight'' (2009), asserts that ``an unpolluted night sky that allows the enjoyment and contemplation of the firmament should be considered a fundamental socio-cultural and environmental right''. The IAU Centre for the Protection of Dark and Quiet Sky from Satellite Interference  (IAU CPS),  co-hosted by the NSF's NOIRLab (the US center for ground-based optical astronomy), and the SKA Observatory (SKAO) was created to respond to the concerns acutely raised by astronomers after the launch of the first 60 Starlink satellites in May 2019. 
The IAU CPS assessed the impact of satellite constellations on astronomy out from the Dark \& Quiet Skies for Science and Society reports \citep{2020dqs1.rept.....W,2021dqs2.rept.....W}.
As man-made objects like satellites and space debris move across the field of view of any astronomical instrument, the astronomical exposure causes streaks across the images and interferes with astronomical data. This is a problem that will be exacerbated for astronomical observatories and many science projects based on wide field instruments and requiring larger time exposures \citep{2022NatAs...6..428L,2020dqs1.rept.....W,2021BAAS...53b0205H,2021zndo...5608826R}.
For instance, the Zwicky Transient Facility has already seen an increase in affected images from 0.5\% in late 2019 to 18\% in August 2021 \citep{2022ApJ...924L..30M}.
The 3.5 degrees wide-field imager of the near-future Vera C. Rubin Observatory in Chile will contain at least one streak in the majority of exposures, further complicated by cascading electronics effects \citep{2020AJ....160..226T}.
SKAO Simulations antenna taking into account main beams and sidelobes pick up from undesired satellite beacons and radio relay data emissions back and forth with ground stations suggest that, once the mature Starlink population is in orbit, every observation in the relevant bands will, on average, take 70\% longer\footnote{\href{https://www.nsf.gov/news/special_reports/jasonreportconstellations/}
{https://www.nsf.gov/news/special\underline{ }reports/jasonreportconstellations}} \citep{2022NatAs...6..428L,lawrencebook}.

Several studies have been used to model the impact of satellite constellations in wide-field instruments, not only for Optical and Infrared Astronomy \citep{2022A&A...657A..75B,2021arXiv211109735M,2022arXiv220912060M}, but also
in Radio Astronomy HI Intensity mapping \citep{2018MNRAS.479.2024H}.
Projects and tools such as the Astrianet telescope sensor network\footnote{\href{https://doi.org/10.18738/T8/GV0ASD}
{https://doi.org/10.18738/T8/GV0ASD}}
provide important information on LEO objects properties and enable further modeling of satellite or debris attitude patterns. This is particularly important to assess impacting brief bright flashes that can occur when a facet or particularly reflective surface of a satellite briefly reflects more sunlight to an observer on the ground \citep{2021AcAau.187..115K}.
Near-future data fusion concepts using space radar and very wide-field tracking telescopes for Space Surveillance and Tracking (SST), such as the PASO telescopes \citep{2022arXiv221104443C},
will provide additional information that is much needed to accurately model the orbital parameters and attitudes of satellites and debris.


Information on the characteristics of a satellite, including space debris, can be obtained from its dynamics, such as its motion around Earth and rotation, or from fusion data with light curve \citep{2007amos.confE..49J,doi:10.2514/1.62986}.
Accurate knowledge of the specific parameters of a satellite in mathematical models is crucial for predicting its motion. Numerical optimization can be used to determine the values of these parameters that best fit both the mathematical models and external data obtained from observations.
To achieve this, an algorithm is created to find the best solution while adhering to the model and given constraints
by iteratively adjusting the decision variables until the objective function is minimized or maximized.
In general, optimization processes may be derived from the gradient method, both for linear systems like the conjugate gradient \citep[e.g.][]{hestenes1952methods} or for nonlinear systems like the descent gradient \citep[e.g.][]{10.2307/43633461}. 
The nonlinear least squares (NLLS) is a well-known gradient based optimization procedure \citep[e.g.][]{doi:10.1137/1.9781611971217}, which suitable addresses a large range of astrodynamics problems \citep{TAPLEY2004xi}. 
Automatic differentiation \citep{10.1145/355586.364792,10.1145/355586.364791,griewank1989automatic} facilitates the evaluations of the derivatives inherent to the NLLS, increasing its computational efficiency to solve astrodynamics problems \citep{10.1007/s40295-023-00378-8}.
On the other hand, non-analytic methods can also be used to address astrodynamics problems. Such kind of optimization procedures do not require the evaluations of derivatives, which may be advantageous to lower computational consumption and / or to be applied to non derivable dynamical models.
For instance, the heuristic optimization procedures Simulated Annealing \citep[e.g.][]{SZU1987157} and Genetic Algorithm \citep[e.g.][]{10.7551/mitpress/3927.001.0001}
have also been proven useful to solve astrodynamics problems \citep[][]{s19040765,gasa2}.
The success of the optimization process depends on the choice of the optimization algorithm, the analytical model, and the quality of the input data, among others factors.

In this paper, two new applications of two different optimization methods are proposed to obtain information on the characteristics of a satellite (or space debris) from observational data. The first method takes advantage of the Theory of Functional Connections (TFC) \citep{U-TFC}, and the second method uses a numerical optimization process through the Nelder-Mead (NM) algorithm \citep{10.1093/comjnl/7.4.308}.

The first method uses the recently developed mathematical framework named the Theory of Functional Connections to solve constrained problems with high efficiency \citep{M-ToC}. The TFC derives constrained expressions, which are a class of functionals that satisfy linear constraints \citep{TFC-multi-tpa}. These constrained expressions are used as functional interpolation to transform a set of nonlinear differential equations subject to constraints into a new set of nonlinear differential equations with no constraints \citep{NDE}. The resulting set of unconstrained nonlinear differential equations can be solved using traditional optimization methods, such as the nonlinear least squares method adopted in this paper.
In the application of the TFC presented here, it is proposed to embed the observational data into the dynamical model.
Thus, any mathematical solution to the dynamical model analytically satisfies the observational data. Note that the solution includes not only the position as a function of time, but also parameters of the dynamics directly related to the properties of the satellite. This is an important state-of-the-art proposed application of the theory, since it allows for a numerical search (based in the dynamical models) of the parameters in a subspace that satisfies the observational data.
Although TFC has been applied in astrodynamics to compute periodic orbits \citep{math9111210,DEALMEIDA2023102068} and to evaluate the costs of transfers between the Earth and Moon \citep{fastTFC} and other systems \citep{allansr}, it has not yet been used to solve the satellite characterization problem.

The second method employed in this study involves using a numerical optimization process based on the Nelder-Mead algorithm. This algorithm was first developed in the 1960s \citep{10.1093/comjnl/7.4.308} and is still being widely used today \citep[e.g.][]{scipy}. The NM algorithm is a heuristic and empirical procedure that operates on a structure called a simplex  using various operations such as ordering, reflection, expansion, contraction, and shrinkage in order to minimize a given function \citep{doi:10.1080/00401706.1962.10490033}. In the context of this study, the NM algorithm is used to minimize the differences between the observational data and the theoretical predictions obtained from the area-to-mass ratio model. This is achieved through iterative operations of the NM algorithm, which result in the values of the model parameters that best fit the observational data. The optimization process is, for instance, carried out using the optimization functions provided by the SciPy library \citep{scipy}, a popular tool for scientific computing in Python, which includes a range of optimization algorithms, including the NM algorithm.



The methodology employed in this research aims to demonstrate and analyze the feasibility of applications of the TFC and NM optimization procedures for satellite characterization. The approach involves simulating observational data and extracting satellite dynamics parameters from these simulated observations. To achieve this, a mathematical model describing the satellite dynamics is utilized within the optimization procedure.
Firstly, the observational data is simulated using both the aforementioned mathematical model (employed in the TFC and NM optimization procedures) and an external, well-known, independent source of data. The methodology and mathematical tools are elaborated in Section \ref{sec:math}, which includes a brief explanation of the TFC and NM methods. Furthermore, the mathematical model utilized to describe the satellite dynamics around the Earth is presented.
In this study, only the two most significant specific forces, namely the effects resulting from the geopotential and drag, are considered. This simplification helps in visualizing the results and analyzing the primary source of errors. It is important to note that this approach does not compromise generality, as similar behavior is expected for the efficiency of both the TFC and NM methods when accounting for other perturbations such as third-body effects and solar radiation pressure, among others.
Section \ref{sec:res} examines the codes based on the presented TFC and NM methods, evaluating their ability to determine the satellite's area-to-mass parameter using a given dataset. A comparison between the TFC and NM methods is conducted in terms of efficiency, encompassing accuracy and resource consumption.
Lastly, the conclusions derived from the research are presented in Section \ref{sec:con}.


\section{Mathematical Tools and Methods}
\label{sec:math}


In this section, the procedures for obtaining parameters related to the dynamics of a satellite are presented. These procedures are based on the application of the TFC and NM optimization methods to satellite characterization. The methods rely on a mathematical model for the dynamics and utilize observational data. 
The purpose of the application of the TFC is to embed the observational data into a new dynamical model, whose mathematical form is to be derived by the theory.
One key characteristic of this method is that any potential solution satisfies the observational data analytically. On the other hand, the proposition based on the NM method offers the advantage of not requiring the evaluation of the gradient. This means that abrupt changes in the trajectory, resulting from one or several impulses, or abrupt variations in the satellite's parameters can be handled. 



\subsection{The Theory of Functional Connections}

Optimization methods, in general, use optimization procedures to satisfy both the differential equations and the constraints. In contrast, optimization methods combined with TFC search for solutions in a subspace that always satisfies constraints analytically \citep{U-TFC}. In this paper, it is proposed that the constraints of the boundary value problem are given by the observational data, which 
are then embed into the dynamical model (equations of motion) by following the outlined procedures. Any subsequent numerical optimization procedure would then search for solutions in a subspace that satisfies the observational data analytically.

The system described by the independent variable $t$ and dependent variable $x(t)$ is subject to a set of $n$ linear constraints on $x$ and its derivative $\dot{x}$.
The general equation to obtain the \textit{constrained expression} involves a free function $g(t)$, a set of $n$ linearly independent support functions $s_k(t)$, and unknown functional coefficients $\eta_k(t,g(t))$, according to \cite{M-ToC}
\begin{eqnarray}\label{eq:ce0}
	x (t, g (t)) = g (t) + \ds\sum_{k = 1}^n \eta_k (t, g (t)) \, s_k (t).
\end{eqnarray}
Using a given set of support functions, the functional coefficients are determined from the constraints, and the constrained expression is obtained. For example, when there is only one constraint for the initial time $t_0$ given by $x(t_0)=x_0$, the functional coefficient can be obtained from Eq.(\ref{eq:ce0}) as $\eta_1 (t, g (t)) = (x_0 - g (t_0))/ s_k (t_0)$. Using the functional coefficient into Eq.(\ref{eq:ce0}), the \textit{constrained expression} is derived as $x (t, g (t)) = g (t) + (x_0 - g (t_0))  s_k (t) / s_k (t_0)$.

If the constrained expression is used in an ordinary differential equation (ODE) given by $\ddot{x} = f(\lambda,\dot{x},x,t)$, where $\lambda$ is a set of $i+1$ unknown coefficients ($\lambda=\lambda_0,\lambda_1,...,\lambda_i$), then it becomes $\ddot{g} = f'(\lambda,\dot{g},g,t)$, where $f'(\lambda,\dot{g},g,t)=f(\lambda,\dot{x (t, g (t))},x (t, g (t)),t)$, and $x(t,g(t))$ is given by the \textit{constrained expression}. It is important to note that there is no constraints in the resulting ODE.

In order to numerically solve the ODE, the free function is expressed as a linear combination of given orthogonal polynomials according to $g (t) = \ds\Sigma_{j = 0}^m \xi_j \, h_j (t) $ truncated at the $m$-th term, where $\xi_j$ are unknown coefficients. After discretizing the resulting ODE for a set of $N$ values of time $t$, the nonlinear least squares optimization method is used to obtain the $[(m+1)+(i+1)]$ unknown coefficients from the $N$ equations generated by the procedure, with $N\ge[(m+1)+(i+1)]$.

\subsection{The Nelder-Mead optimization procedure}

The Nelder-Mead optimization procedure is a popular iterative algorithm used to minimize an objective function of several variables \citep{10.1093/comjnl/7.4.308} - see also \citep{Press2007,scipy}. It is considered a direct search algorithm because it does not require the computation of gradient information, making it useful in cases where the objective function is non-differentiable or when the gradient is too difficult to compute.

The algorithm begins by selecting an initial set of points, called a simplex, to cover a wide range of the parameter space. For example, consider an objective function with two variables. The simplex would consist of three points, 
forming a triangle in the parameter space. The algorithm then evaluates the objective function at each of the simplex points and identifies the best point with the smallest function value.

The point with the highest objective function value is then reflected through the centroid of the remaining points in the simplex. This reflected point is compared with the other points in the simplex. If the reflected point has a lower objective function value than the lowest point in the simplex, the reflected point is expanded further in the same direction to create a new point. This new point is then evaluated to determine if it is better than the previous simplex points. If the reflected point does not improve the objective function value, the simplex is contracted towards the lowest point. This contraction reduces the size of the simplex in the direction of the highest point, creating a smaller simplex that is closer to the minimum point.

The algorithm continues to iterate through these steps until a stopping criterion is met. This criterion may be a maximum number of iterations or a change small enough in the objective function value, which is defined by a value for the tolerance to satisfy the minimization procedure. The Nelder-Mead algorithm is relatively simple to implement and has been shown to be effective in a wide range of optimization problems. However, it may not converge to the global minimum of the objective function and may get stuck in local minima. It is also sensitive to the choice of the initial simplex and the parameters that control the size of the operations on the simplex. Therefore, careful tuning of these parameters may be necessary to achieve good performance.

The Nelder-Mead method is a gradient-free optimization algorithm suitable for problems with a moderate number of parameters and noisy objective functions. It is effective for non-linear, non-convex, and complex objective functions. The method is simple to implement and computationally efficient as it does not require gradient or Hessian calculations. Other optimization algorithms such as Powell's method, COBYLA, and Basin Hopping can also provide good results, but comparing these methods is not the focus of this work.



\subsection{Dynamical model for satellite characterization}

The dynamics of a satellite in orbit around the Earth can be described in the International Celestial Reference Frame (ICRF) as
\begin{eqnarray}\label{eq:em}
	\ddot{\B{r}} = \B{g}_{10}-\frac{1}{2}C_D\frac{ A}{m} \rho v_r \B{v}_r+\B{P}.
\end{eqnarray}
In the above equation, $\B{r}$ denotes the position of the satellite in the ICRF, and the dots indicate derivatives with respect to time. The term $\B{g}_{10}$ represents the gravitational interaction with the Earth, which is modeled using the EGM96 model, truncated in the degree and order 10. Its calculation is performed using the algorithm outlined in \cite{kugaegm}, along with another algorithm proposed in this paper. The purpose of incorporating the additional algorithm is to reduce the computational resources required for implementing the TFC method code, as presented in \ref{app:A}.
The area-to-mass ratio $A/m$ is the transverse area of the satellite divided by its mass. The coefficient of drag $C_D$ is a measurement of the resistance that the satellite experiences due to the atmosphere (the usual value $C_D=2.2$ is used \citep{COOK1965929,gaposchkin1988analysis}) and $\rho (\B{r},t)$ is the density of the atmosphere, which is a function of both position and time according to the Jacchia-Roberts atmospheric dynamic model \citep{Jacchia_1964, Jacchia_1965, roberts_1971}. The variable $\B{v}_r$, with magnitude $v_r$, is the velocity of the satellite relative to the International Terrestrial Reference Frame (ITRF), which rotates with the Earth and is connected to the ICRF through the Earth Orientation Parameters \citep{eop}. The term $\B{P}$ represents the other perturbative specific forces acting on the satellite, such as third-body perturbations, solar radiation pressure, etc. In order to simplify and enhance comprehension of the outcomes obtained from the application of the two approaches proposed in this study, the term represented by $\B{P}$ is not considered when generating the results of this paper. 



\subsection{Source of the simulated observational data}

In this paper, the TFC and NM methods are used to estimate the area-to-mass parameter using simulated observational data. It is important to emphasize that the application of the TFC and NM optimization procedures is based on Eq.~(\ref{eq:em}). Therefore, if the simulated observational data is generated also using the mathematical model described by Eq.~(\ref{eq:em}), it is expected that the methods will retrieve the area-to-mass parameter with high precision. This scenario represents an ideal situation where the observational data perfectly aligns analytically with the model.
However, it is worth noting that this ideal situation does not represent real scenarios, where perturbation forces act on the satellite and the observational data cannot be analytically described by the available mathematical model. Consequently, the TFC and NM methods are also applied in situations where the simulated observational data is generated independently, rather than using the model described by Eq.~(\ref{eq:em}). In this case, the General Mission Analysis Tool\footnote{The GMAT software is developed and maintained by NASA's Goddard Space Flight Center to model, simulate, and optimize spacecraft trajectories from low Earth orbits to interplanetary missions. \href{https://software.nasa.gov/software/GSC-17177-1}{https://software.nasa.gov/software/GSC-17177-1}} (GMAT) is used to simulate the observational data. Although efforts are made to adjust the GMAT software to generate results as close as possible to the mathematical model shown in Eq.~(\ref{eq:em}), the results obtained do not perfectly match due to the complexity of the algorithms involved and the inherent differences between the mathematical models.
Thus, in this paper, the observational data is simulated using both \textit{our data generator} - based on the dynamical model given by Eq.~(\ref{eq:em}) - and GMAT as reference solutions for comparison. The optimization process with the generation of simulated observational data using GMAT is described below:
\begin{enumerate}
    \item An initial (original) value for the area-to-mass ratio is given.
    \item GMAT is employed with the initial area-to-mass ratio and a set of initial conditions to generate a set of boundary conditions.
    \item The set of boundary conditions generated in the previous step is used as simulated observational data in the TFC and NM optimization processes to estimate (retrieve) the area-to-mass ratio.
    \item The estimated (retrieved) area-to-mass ratio is compared with the initial value given in step 1.
\end{enumerate}
In the case where the observational data is simulated using \textit{our data generator}, the process is similar to the one described above, except that, in step 2, \textit{our data generator} (based on Eq. \ref{eq:em}) is used to generate the set of boundary conditions, instead of GMAT.

\section{Results}
\label{sec:res}

In this paper, the TFC and Nelder-Mead optimization procedures are used to determine the parameters of a spacecraft using available position and velocity data. Specifically, the area-to-mass ratio is estimated using position and velocity data at two different times. These required data are generated by integrating initial conditions with a specific value of the area-to-mass ratio for a given time period, T.
Hence, the inputs for the implemented methods are the initial position and velocity (at $t_0=$ ``03 Feb 2023 00:00:00 (UTC)'') and the final position and velocity (at $t_f=t_0+T$, with $T=3$ minutes), while the output is the area-to-mass ratio.

The original parameter refers to the value used to simulate the observational data (using GMAT software or our data generator), while the retrieved parameter denotes the value obtained by the TFC and NM methods from the simulated observational data. The disparities between the retrieved area-to-mass ratio and the original value are analyzed in this section with respect to altitude, latitude, longitude, and the area-to-mass ratio for both TFC and NM methods. This difference is better visualized using a relative error index, defined as 
\begin{eqnarray}\label{eq:re}
\text{relative error} = \bigg|\frac{{(A/m)}_r-{(A/m)}_o}{{(A/m)}_o}\bigg|,
\end{eqnarray}
where ${(A/m)}_r$ represents the retrieved area-to-mass ratio and ${(A/m)}_o$ denotes the original area-to-mass ratio used to generate the data.

It is important to note that altitudes ranging from 150 km to 800 km are considered a critical range where the atmosphere's effect can significantly influence the orbit within a short time \citep{ASHENBERG199675}. The magnitude of this effect varies depending on the altitude; satellites below 300 km experience a 12 km decrease in altitude per day, while those around 800 km may encounter a similar decrease in altitude for over 20 days \citep{gaposchkin1988analysis}. To evaluate the methods' performance, orbits with altitudes ranging from 150 km to 800 km are chosen, with a step size of 50 km, for various latitude and longitude values. No problem of convergence for the initial guess has been observed for both methods.

\subsection{Accuracy of the retrieved parameter as function of the altitude}
The relative errors (Eq.~(\ref{eq:re}))
are shown in Table \ref{tab:am1} for the original area-to-mass parameters $A/m=1$ m$^2$/kg and $A/m=0.01$ m$^2$/kg. The TFC and NM methods are compared against the data generated using GMAT and \textit{our data generator}, with latitude and longitude equal to zero. As altitude increases, the density of the atmosphere decreases, and then, also the drag perturbation, which makes the optimization process of a parameter proportional to the drag more difficult, and leads to an increase in the relative error shown in Table \ref{tab:am1}. Despite using a simple integration algorithm based on a simplified Bulirsch-Stoer method in the Nelder-Mead method or minimizing residuals through the nonlinear least squares method in the TFC method, the retrieved parameters are reasonably close to the original values used by GMAT, showing a maximum relative error of about 5\% and 2\% for extreme cases of high altitudes (close to 800 km) and a small area-to-mass ratio (0.01 m$^2$/kg) for the TFC and NM methods, respectively. Table \ref{tab:am1} shows that the relative error is up to the order of $10^{-4}$ for altitudes up to 450 km and $10^{-6}$ for a higher area-to-mass (1 m$^2$/kg) for both the TFC and NM methods. However, when the observational data is simulated using \textit{our data generator}, the relative error is much smaller. Its order of magnitude varies from $10^{-12}$ to $10^{-7}$ for the TFC method and for the original area-to-mass parameter ($A/m=0.01$ m$^2$/kg). The relative error decreases by two orders of magnitude when the original parameter is $A/m=1$ m$^2$/kg. The relative error for the NM method is about two orders of magnitude higher compared to the TFC method when the parameter is retrieved from observational data simulated using \textit{our data generator}.
\begin{table*}[ht]
	\caption{Relative error between the retrieved and original values of the area-to-mass parameter (Eq.~\ref{eq:re}) obtained using the TFC and NM methods from data of position and velocity evaluated at two times with difference of 3 minutes simulated for $A/m$=0.01 m$^2$/kg and $A/m$=1 m$^2$/kg.}
	\centering
	\resizebox{1.2\textwidth}{!}
    {
		\begin{tabular}{|c|c|c|c|c|c|c|c|c|}
            \hline
            &\multicolumn{8}{c|}{Relative error}\rule[-7pt]{0pt}{17pt}\\[0.ex]
            \cline{2-9}
			&\multicolumn{4}{c|}{TFC method}&\multicolumn{4}{c|}{NM method}\rule[-7pt]{0pt}{17pt}\\[0.ex]
            \cline{2-9}
            Altitude&\multicolumn{4}{c|}{Parameter retrieved from observational simulated data generated by }&\multicolumn{4}{c|}{Parameter retrieved from observational simulated data generated by}\rule[-7pt]{0pt}{17pt}\\[0.ex]
            \cline{2-9}
           (km)&\multicolumn{2}{c|}{GMAT}&\multicolumn{2}{c|}{\textit{our data generator}}&\multicolumn{2}{c|}{GMAT}&\multicolumn{2}{c|}{\textit{our data generator}}\rule[-7pt]{0pt}{17pt}\\[0.ex]
           \cline{2-9}
           &$A/m$=0.01 m$^2$/kg&$A/m$=1 m$^2$/kg&$A/m$=0.01 m$^2$/kg&$A/m$=1 m$^2$/kg&$A/m$=0.01 m$^2$/kg&$A/m$=1 m$^2$/kg&$A/m$=0.01 m$^2$/kg&$A/m$=1 m$^2$/kg\rule[-7pt]{0pt}{17pt}\\[-1.ex]
   \hline
   \hline
150&$3.39\times 10^{-07}$&$3.28\times 10^{-07}$&$7.41\times10^{-12}$&$6.91\times10^{-14}$ &5.96$\times 10^{-08}$&3.22$\times 10^{-07}$ &1.51$\times 10^{-09}$&4.66$\times 10^{-11}$ \\
200&$4.80\times 10^{-05}$&$5.28\times 10^{-05}$&$5.59\times10^{-12}$&$1.08\times10^{-13}$ &1.08$\times 10^{-04}$&1.09$\times 10^{-04}$ &1.07$\times 10^{-08}$&7.57$\times 10^{-11}$ \\
250&$1.73\times 10^{-05}$&$1.64\times 10^{-07}$&$1.27\times10^{-10}$&$2.18\times10^{-12}$ &6.20$\times 10^{-06}$&2.79$\times 10^{-07}$ &4.74$\times 10^{-08}$&4.66$\times 10^{-11}$ \\
300&$4.87\times 10^{-05}$&$1.44\times 10^{-07}$&$5.94\times10^{-10}$&$6.94\times10^{-12}$ &2.18$\times 10^{-05}$&1.46$\times 10^{-07}$ &8.41$\times 10^{-08}$&2.25$\times 10^{-09}$  \\
350&$1.19\times 10^{-04}$&$8.50\times 10^{-07}$&$1.23\times10^{-09} $&$6.42\times10^{-12}$ &4.69$\times 10^{-05}$&1.66$\times 10^{-07}$ &1.57$\times 10^{-07}$&1.02$\times 10^{-09}$  \\
400&$2.73\times 10^{-04}$&$2.38\times 10^{-06}$&$1.69\times10^{-09} $&$2.47\times10^{-12}$ &1.17$\times 10^{-04}$&6.91$\times 10^{-07}$ &3.32$\times 10^{-07}$&1.03$\times 10^{-08}$ \\
450&$5.97\times 10^{-04}$&$5.61\times 10^{-06}$&$6.83\times10^{-09} $&$6.09\times10^{-11}$ &2.57$\times 10^{-04}$&2.19$\times 10^{-06}$ &7.23$\times 10^{-07}$&1.03$\times 10^{-08}$  \\
500&$1.26\times 10^{-03}$&$1.22\times 10^{-05}$&$2.65\times10^{-08} $&$1.14\times10^{-10}$ &5.04$\times 10^{-04}$&5.13$\times 10^{-06}$ &7.23$\times 10^{-07}$&2.98$\times 10^{-09}$  \\
550&$2.60\times 10^{-03}$&$2.56\times 10^{-05}$&$1.69\times10^{-10}$&$6.48\times10^{-10}$ &1.01$\times 10^{-03}$&1.22$\times 10^{-05}$ &1.48$\times 10^{-07}$&8.31$\times 10^{-08}$  \\
600&$5.23\times 10^{-03}$&$5.19\times 10^{-05}$&$6.11\times10^{-09} $&$3.30\times10^{-10}$ &2.46$\times 10^{-03}$&2.80$\times 10^{-05}$ &4.74$\times 10^{-08}$&7.62$\times 10^{-08}$  \\
650&$1.01\times 10^{-02}$&$1.01\times 10^{-04}$&$1.47\times10^{-08} $&$7.89\times10^{-10}$ &4.75$\times 10^{-03}$&4.43$\times 10^{-05}$ &1.53$\times 10^{-05}$&4.41$\times 10^{-07}$  \\
700&$1.90\times 10^{-02}$&$1.90\times 10^{-04}$&$4.66\times10^{-08} $&$1.02\times10^{-10}$ &8.17$\times 10^{-03}$&8.84$\times 10^{-05}$ &5.62$\times 10^{-05}$&1.07$\times 10^{-06}$ \\
750&$3.39\times 10^{-02}$&$3.38\times 10^{-04}$&$3.31\times10^{-07} $&$2.36\times10^{-09} $ &1.32$\times 10^{-02}$&1.15$\times 10^{-04}$ &1.07$\times 10^{-08}$&5.66$\times 10^{-07}$ \\
800&$5.65\times 10^{-02}$&$5.65\times 10^{-04}$&$1.13\times10^{-07} $&$5.04\times10^{-09} $ &1.98$\times 10^{-02}$&2.36$\times 10^{-04}$ &1.31$\times 10^{-04}$&4.07$\times 10^{-06}$  \\\hline
\end{tabular}
	}
	\label{tab:am1}
\end{table*}

\subsection{Accuracy for several values of latitude and longitude}

\begin{figure*}[!t]
	\centering
    \includegraphics[width=0.88\linewidth]{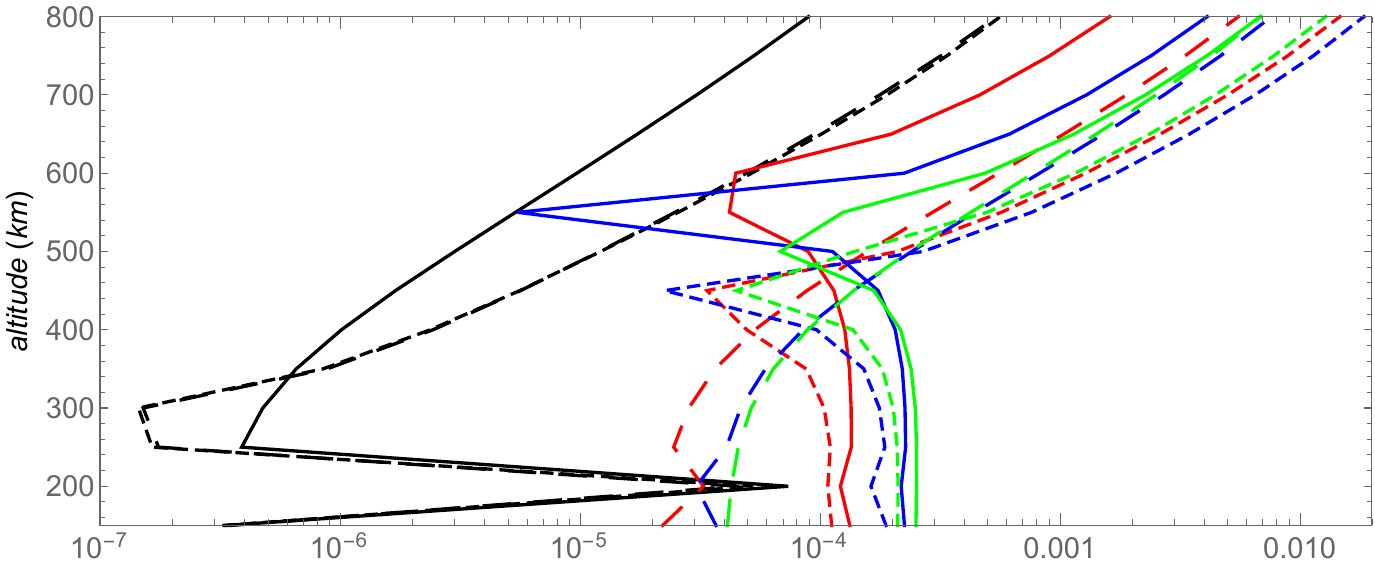}
    \includegraphics[width=1.\linewidth]{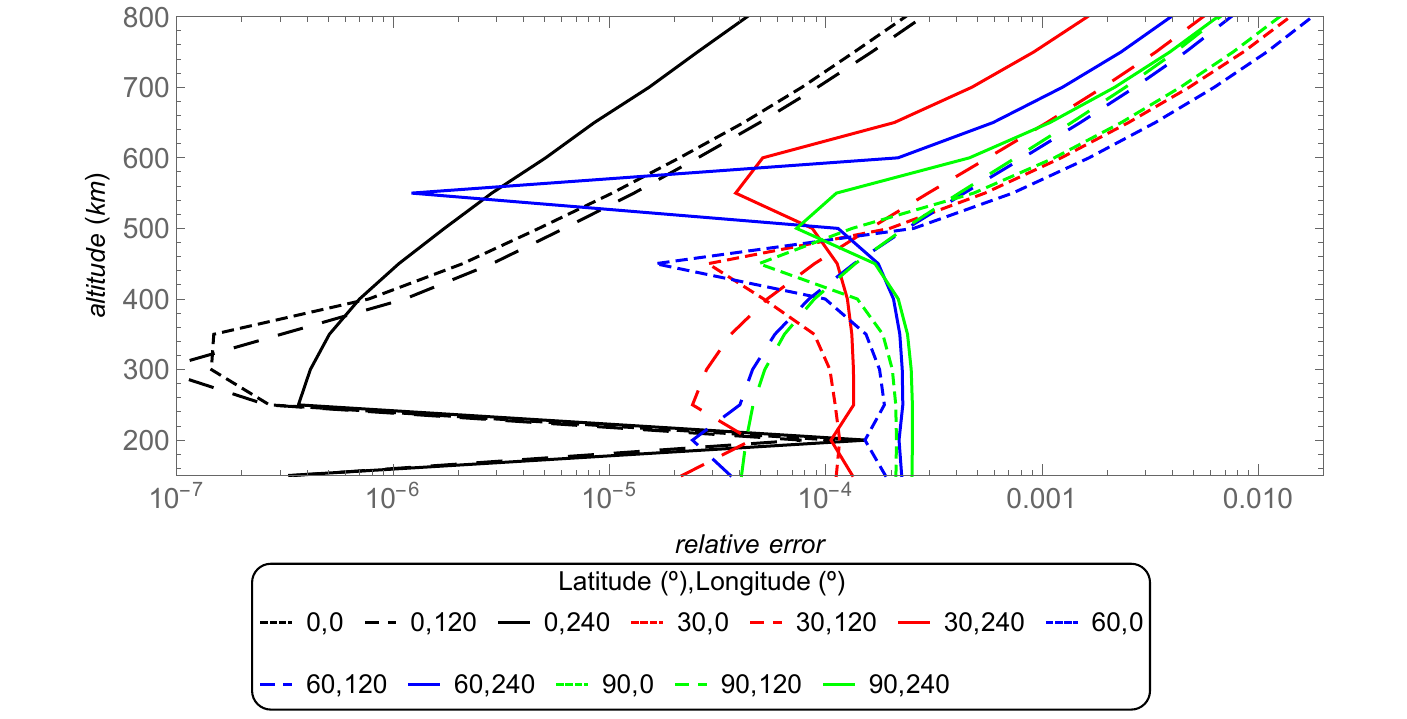}
	\caption{Relative error of the parameter retrieved using the TFC (above) and the NM (below) methods as function of the altitude for the area-to-mass ratio $A/m=1$ m$^2$/kg. The observational data are simulated using GMAT.}
	\label{fig:diag2_re}
\end{figure*}

Figure \ref{fig:diag2_re} displays the altitude-dependent relative error obtained using the TFC method (top panel) and NM method (bottom panel) with various initial latitudes and longitudes. The data was generated through the GMAT software under the condition of $A/m=1$ m$^2$/kg, and the velocity was selected such that the initial orbit's eccentricity is zero, while its direction depends on the longitude. The TFC and NM methods yield a relative error below 0.01 for altitudes up to roughly 700 km, with the relative error order of magnitude ranging from $10^{-7}$ to $10^{-3}$ in this range. However, for altitudes exceeding 500 km, the relative error increases.

\begin{figure*}[!t]
	\centering
    \includegraphics[width=0.89\linewidth]{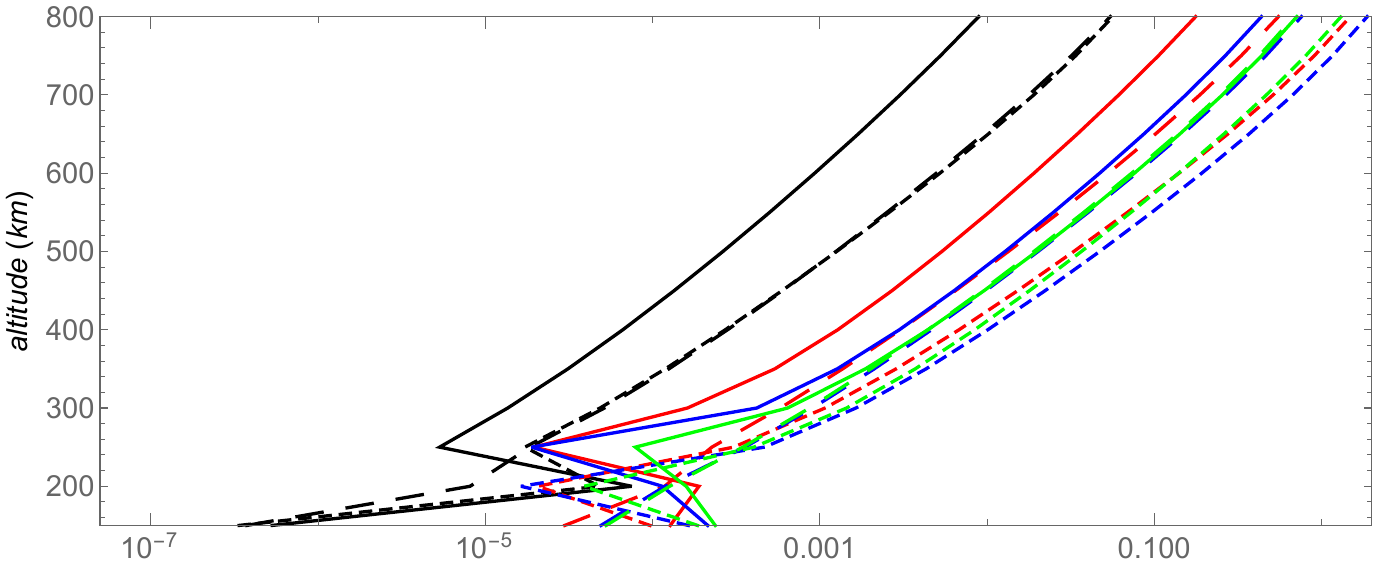}
    \includegraphics[width=1\linewidth]{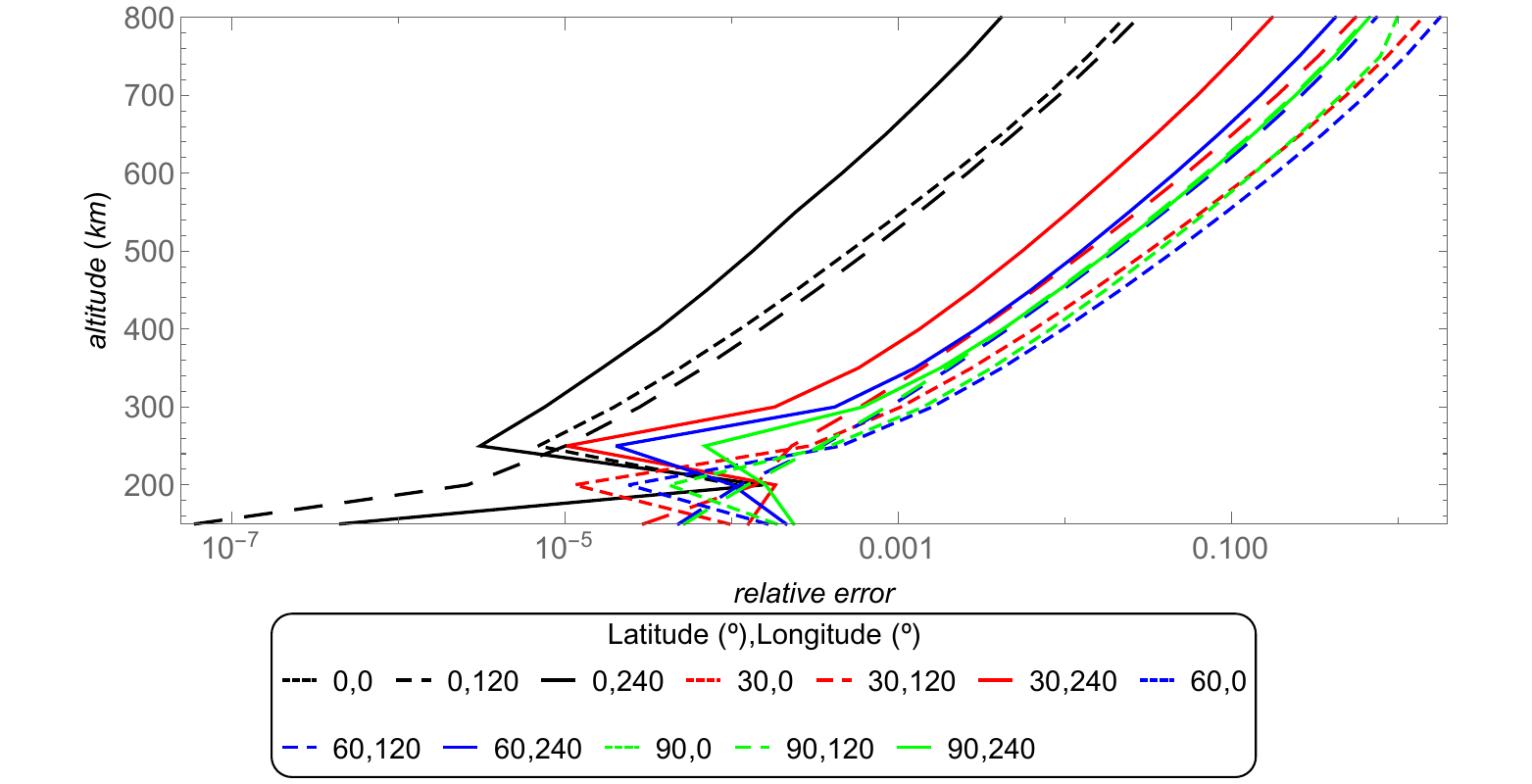}
	\caption{Relative error of the parameter retrieved using the TFC (above) and the NM (below) methods as function of the altitude for the area-to-mass ratio $A/m=0.01$ m$^2$/kg. The observational data are simulated using GMAT.}
	\label{fig:diag2nm}
\end{figure*}

Figure \ref{fig:diag2nm} illustrates the argument for an area-to-mass ratio ($A/m$) of 0.01 m$^2$/kg. The TFC and Nelder-Mead methods produce comparable outcomes. However, the precision of the recovered parameter's relative error increases at altitudes greater than 250 km for both methods. Again, the relative error rises at 200 km altitude. Notably, the relative error exhibits a two-order-of-magnitude increase compared to the results showed in Figure \ref{fig:diag2_re} for $A/m$ = 1 m$^2$/kg.

Overall, the relative error increases significantly for altitudes exceeding 500 km when $A/m$=1 m$^2$/kg or 250 km when $A/m$=0.01 m$^2$/kg. Trajectories with low area-to-mass ratios resemble those with zero drag, particularly at high altitudes. Although differences in relative error exist depending on latitude or longitude, the trend of increasing error with altitude remains consistent globally.

\subsection{Accuracy for observational data simulated using \textit{our data generator}}

\begin{figure*}
	\centering
    \includegraphics[width=0.84\linewidth]{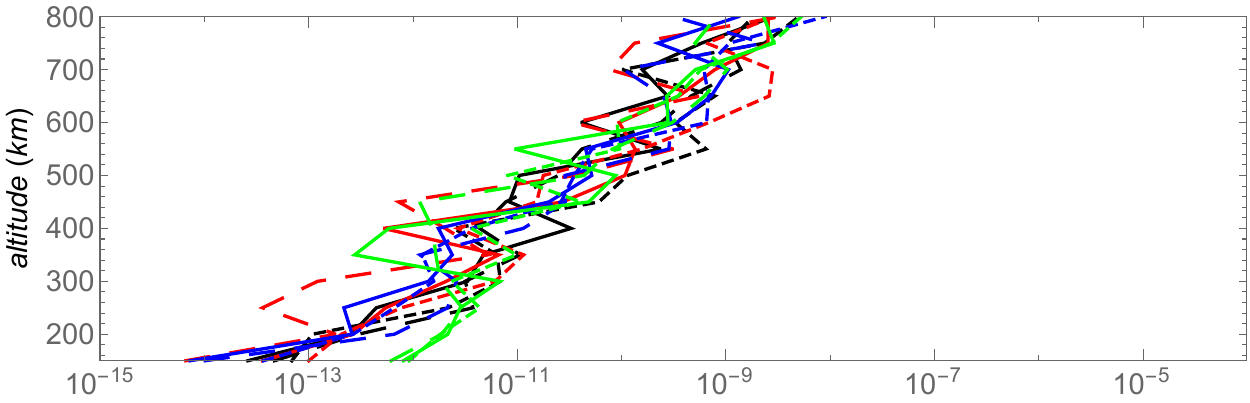}
    \includegraphics[width=1.\linewidth]{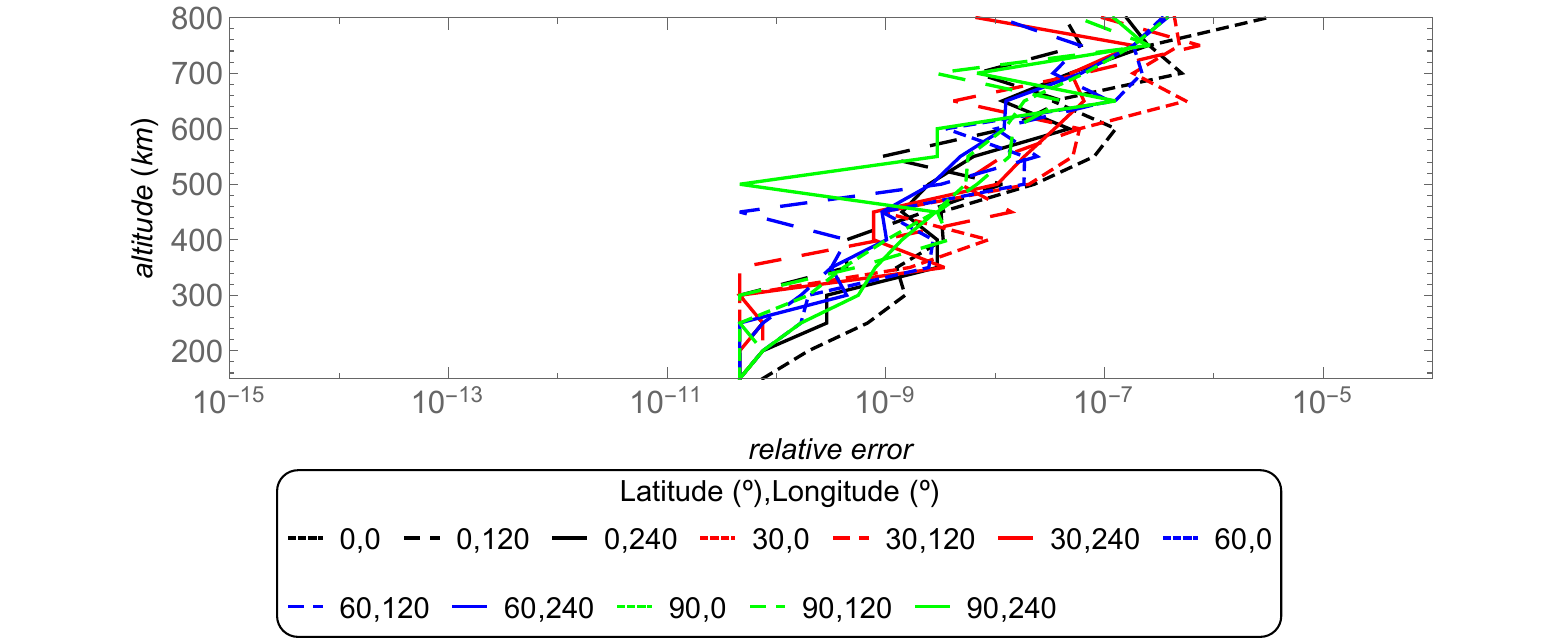}
	\caption{Relative error as function of the altitude for an area-to-mass ratio $A/m=1$ m$^2$/kg evaluated through the TFC (above) and NM (below) methods. The data is simulated using \textit{our data generator}.}
	\label{fig:diag3}
\end{figure*}

\begin{figure*}
	\centering
    \includegraphics[width=0.8\linewidth]{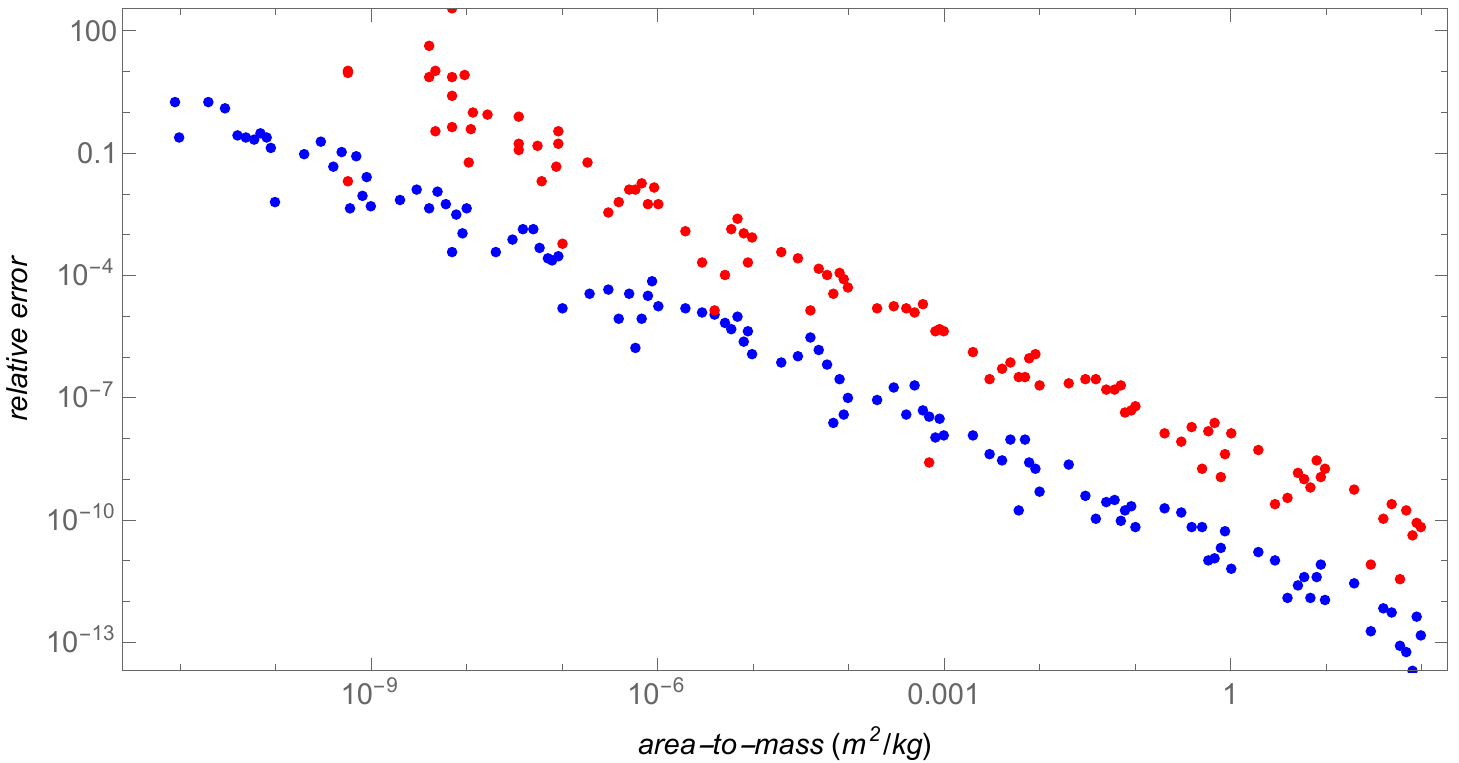}
	\caption{Relative error as function of the area-to-mass ratio using the TFC (blue) and the NM (red) methods for an orbit with altitude of 400 km. The data is generated using \textit{our data generator}.}
	\label{fig:diag4}
\end{figure*}

The results for observational data simulated using \textit{our data generator} are presented in Figure \ref{fig:diag3} for an area-to-mass ratio of $A/m=1~\text{m}^2/\text{kg}$. For the TFC method, the order of magnitude of the relative error varies from $10^{-14}$ for lower altitudes to $10^{-9}$ for orbits with altitudes close to 800 km. In comparison, the relative error of the retrieved parameter using the NM method is a few orders of magnitude higher, as can be seen in Figure \ref{fig:diag3}. Note that the relative errors are much lower compared to the data generated by GMAT, as expected. This figure illustrates the accuracy limits of the methods in the ideal scenario where the mathematical models describe the data with the highest possible accuracy (machine-level accuracy). It is important to note that even in this case, the models are not simple, and there are multiple steps of complex numerical evaluations required to obtain the optimized parameter's final value. The relative error as a function of the area-to-mass ratio is shown in Figure \ref{fig:diag4}, for the case where the initial position has an altitude and longitude equal to zero and the altitude is 400 km for both TFC (blue) and NM (red) methods using observational data simulated by \textit{our data generator}, i.e., a similar model.


\subsection{Influence of the quality of the observational data over the retrieved parameter}

This section aims to assess the impact of data quality on the performance of the proposed methods by incorporating simulated noises. To account for various observation techniques, a random error, denoted as $x$, is introduced to both the position and velocity data. This error follows a Gaussian distribution with a probability function given by
\begin{equation}\label{eq:sig}
	p(x) = \frac{1}{\sqrt{2 \pi \sigma^2}} e^{ - \frac{(x - \mu)^2}{2 \sigma^2}},
\end{equation}
where $\mu$ represents the mean and $\sigma$ denotes the standard deviation of the distribution.
The specific standard deviation value depends on the level of accuracy and the characteristics of the observation system or technique being simulated. It is important to note that in practice, the accuracy of Low Earth Orbit (LEO) objects can vary significantly based on factors such as the instrumentation, tracking method, atmospheric conditions, altitude, hardware limitations, and other factors.

Satellite Laser Ranging (SLR) utilizes laser pulses to measure the range between a ground station and a satellite, providing highly accurate measurements typically within a range of a few centimeters to a few millimeters in terms of range accuracy \citep{Pearlman_2002}. Therefore, for simulating SLR observations, a possible standard deviation value could be in the range of 0.01 m to 0.1 m for positions. Considering the uncertainties for positions, it is assumed that the uncertainties for velocity are in the range of 0.001 m/s to 0.01 m/s.

Global Navigation Satellite Systems (GNSS) such as GPS, GLONASS, and Galileo offer precise positioning capabilities. The accuracy of GNSS-based position measurements for LEO satellites can range from one decimeter using GNSS broadcast ephemerides  \citep{navi416} to a few meters or up to tens of meters, depending on factors such as the number and quality of received satellite signals, atmospheric conditions, and receiver hardware \citep{Hofmann2008}. Therefore, for simulating GNSS-based observations, a standard deviation value in the range of 0.1 to 10 meters could be considered for positions. For the velocity, an uncertainty in the range of 0.01 m/s to 1 m/s is assumed.

Furthermore, differences between the positions and velocities obtained with Two-Line Elements (TLEs) when compared with information obtained from GPS is up to 140 meters in position \citep{Chen2017,s22082902} and from 0.1 m/s \citep[pg. 38]{Chen2017} up to 140 m/s \citep{s22082902} in velocity.

By incorporating simulated noises into the positional and velocity data, we can explore the impact of data quality on the performance of the proposed methods and gain insights into their robustness and effectiveness under realistic conditions. To simulate real-world scenarios and observe the influence of the standard deviation $\sigma$ on the retrieved parameter, an initial position with latitude, longitude, and altitude set to 0°, 0°, and 200 km, respectively, is considered.

Table \ref{tab:noise1} shows the median and standard deviation of the retrieved parameter with respect to the original values $A/m = 1 , \text{m}^2/\text{kg}$ and $A/m = 0.01 , \text{m}^2/\text{kg}$ for several combination values of the standard deviation $\sigma$ in the range of $10^{-2}$ m to $10^2$ m and $10^{-3}$ m/s to $10^0$ m/s for position and velocity, respectively.

For SLR observations, with uncertainties in the range of $10^{-2}$ m or $10^{-1}$ m for position and $10^{-2}$ m/s for velocity, Table \ref{tab:noise1} indicates that the standard deviation for the retrieved parameter is on the order of $10^{-3}$ m$^2$/kg.

For GNSS-based observations, with uncertainties in the range of $10^{-1}$ m or $10^0$ m for position and $10^{-2}$ m/s or $10^{-1}$ m/s for velocity, the standard deviation of the retrieved parameter is also on the order of $10^{-3}$ m$^2$/kg when the uncertainty in the initial velocity is $10^{-2}$ m/s. This is sufficient to retrieve the area-to-mass ratio from simulated observational data with a standard deviation lower than the value of the parameter. However, for larger uncertainties in the velocity, on the order of $10^{-1}$ m/s, the standard deviation of the retrieved parameter is on the order of $10^{-2}$ m$^2$/kg, which is lower than the original value of the area-to-mass ratio for $A/m = 1 , \text{m}^2/\text{kg}$ but higher if the original parameter is $A/m = 0.01 , \text{m}^2/\text{kg}$. Thus, the data obtained from GNSS-based observations could still be used to retrieve the parameter for larger area-to-mass ratios using the information given by the assumed constraints (two positions and two velocities).

In comparison with SLR and GNSS-based observations, the uncertainties obtained from TLEs are much higher, typically on the order of $10^{1}$ m for position and from $10^{-1}$ to $10^{0}$ m/s for velocity. In this case, the standard deviation of the retrieved area-to-mass parameter is on the order of $10^{-1}$ m$^2$/kg.

As shown in this subsection, the influence of errors due to the real observations is much larger than the influence of the errors inherent to the procedures of the numerical applications of the proposed methods (both TFC and NM). 
This makes the methods suitable for the purpose of satellite characterization.

\begin{table*}
	\caption{The median and standard deviation of the retrieved parameter with respect to the original area-to-mass $A/m$=1 m$^2$/kg and $A/m$=0.01 m$^2$/kg obtained with the TFC and NM methods from data with noise. The initial position is such that its latitude, longitude, and altitude are 0º, 0º, and 200 km, respectively.}
	\centering
	\resizebox{1.\textwidth}{!}
    {
		\begin{tabular}{|c|c|c|c|c|}
   \hline
            Original&\multicolumn{2}{|c|}{Standard deviation of}&Parameter retrieved from TFC method&Parameter retrieved from NM method\rule[-7pt]{0pt}{17pt}\\[0.ex]
            value&\multicolumn{2}{|c|}{the initial distribution}&&\rule[-7pt]{0pt}{17pt}\\[0.ex]
   \hline
    $A/m$ ($\text{m}^2/\text{kg}$)&position (m)&velocity (m/s)& Median $\pm$ Standard deviation ($\text{m}^2/\text{kg}$)&  Median $\pm$ Standard deviation ($\text{m}^2/\text{kg}$)\rule[-7pt]{0pt}{17pt}\\[0.ex]
   \hline
\hline
&$1\times10^{2}$&$1\times10^{0}$&$0.94\pm0.52$&$0.98\pm0.75$\\
&$1\times10^{1}$&$1\times10^{0}$&$0.94\pm0.41$&$0.99\pm0.56$\\
&$1\times10^{0}$&$1\times10^{0}$&$1.10\pm0.44$&$1.01\pm0.52$\\
1&$1\times10^{0}$&$1\times10^{-1}$&$1.002\pm0.043$&$1.000\pm0.062$\\
&$1\times10^{-1}$&$1\times10^{-1}$&$1.005\pm0.046$&$1.013\pm0.059$\\
&$1\times10^{-1}$&$1\times10^{-2}$&$1.0007\pm0.0045$&$1.0005\pm0.0064$\\
&$1\times10^{-2}$&$1\times10^{-2}$&$0.9994\pm0.0041$&$1.0000\pm0.0062$\\
&$1\times10^{-2}$&$1\times10^{-3}$&$1.00002\pm0.00045$&$1.00004\pm0.00061$\\
\hline
\hline
&$1\times10^{0}$&$1\times10^{-2}$&$(1.06\pm0.42)\times10^{-2}$&$(1.16\pm0.74)\times10^{-2}$\\
0.01&$1\times10^{-1}$&$1\times10^{-2}$&$(0.90\pm0.44)\times10^{-2}$&$(0.96\pm0.56)\times10^{-2}$\\
&$1\times10^{-2}$&$1\times10^{-2}$&$(1.00\pm0.47)\times10^{-2}$&$(1.10\pm0.57)\times10^{-2}$\\
&$1\times10^{-2}$&$1\times10^{-3}$&$(1.002\pm0.049)\times10^{-2}$&$(0.995\pm0.059)\times10^{-2}$\\
\hline
\end{tabular}
	}
	\label{tab:noise1}
\end{table*}

\subsection{Simulated observational data given only by positions}

The results of applying the optimization procedures developed in this paper are shown above for the case where there is information on the initial position and velocity, as well as the position and velocity after 3 minutes.

In this subsection, the constraints are limited to four positions evaluated at four different times: 0, T/3, 2T/3, and T. It should be noted that in this case, the codes retrieve the parameter assuming that no information on the velocity is contained in the simulated observational data. The optimization codes estimate the three components of the velocity together with the area-to-mass parameter. This consideration increases the execution time of the Nelder-Mead algorithm, but it does not affect the execution time of the TFC-based method.

Table \ref{tab:noise2} demonstrates that the standard deviation values of the retrieved parameters are similar in magnitude to those shown in Table \ref{tab:noise1} for the case where the standard deviation of velocity (in m/s) is one order of magnitude lower than the standard deviation of position (in m). The retrieved values are one order of magnitude lower than those shown in Table \ref{tab:noise1} when the standard deviation of the initial velocity distribution (in m/s) is of the same magnitude as the standard deviation of the initial position distribution (in m). Hence, if the standard deviation of velocity (in m/s) is of the same magnitude or higher than the standard deviation of position (in m), it is more accurate to retrieve the parameter using only position information.

\begin{table*}
	\caption{The median and standard deviation of the retrieved parameter with respect to the original area-to-mass $A/m$=1 m$^2$/kg and $A/m$=0.01 m$^2$/kg obtained with the TFC and NM methods from data with noise. The simulated observational data are given by 4 values of position (for $t=t_0,t_0+T/3,t_0+2T/3,t_0+T$, with $T=3$ minutes). The initial position is such that its latitude, longitude, and altitude are 0º, 0º, and 200 km, respectively.}
	\centering
	\resizebox{1.\textwidth}{!}
    {
		\begin{tabular}{|c|c|c|c|c|c|c|}
   \hline
            Original&\multicolumn{1}{|c|}{Standard deviation of}&Parameter retrieved from TFC method&Parameter retrieved from NM method\rule[-7pt]{0pt}{17pt}\\[0.ex]
            value&\multicolumn{1}{|c|}{the initial distribution}&&\rule[-7pt]{0pt}{17pt}\\[0.ex]
   \hline
    $A/m$ ($\text{m}^2/\text{kg}$)&position (m)& Median  $\pm$ Standard deviation ($\text{m}^2/\text{kg}$)& Median $\pm$ Standard deviation ($\text{m}^2/\text{kg}$)\rule[-7pt]{0pt}{17pt}\\[0.ex]
   \hline
\hline
&$1\times10^{1}$ &1.01$\pm0.17$&1.00$\pm0.13$\\
1&$1\times10^{0}$&0.999$\pm0.014$&0.999$\pm0.015$\\
&$1\times10^{-1}$&0.9999$\pm0.0016$&1.0002$\pm0.0016$\\
&$1\times10^{-2}$&0.99999$\pm0.00018$&1.00000$\pm0.00071$\\
&$1\times10^{-3}$&1.000000$\pm0.000015$&1.000000$\pm0.000124$\\
\hline
\hline
0.01&$1\times10^{-1}$&$(1.00\pm0.13)\times10^{-2}$&$(1.01\pm0.17)\times10^{-2}$\\
&$1\times10^{-2}$    &$(0.999\pm0.014)\times10^{-2}$&$(1.004\pm0.015)\times10^{-2}$\\
&$1\times10^{-3}$    &$(1.0000\pm0.0014)\times10^{-2}$&$(1.0001\pm0.0015)\times10^{-2}$\\
\hline
\end{tabular}
	}
	\label{tab:noise2}
\end{table*}

\subsection{Execution time}

The results for the TFC method were obtained using the Python programming language with the assistance of the TFC module \citep{tfc2021github} that uses automatic differentiation and a just-in-time (JIT) compiler \citep{JaxGithub}.
On the other hand, the results for the NM method were obtained using the Python programming language with the help of the SciPy library \citep{scipy}.
Since every iteration in the NM procedure depends on the results of the integration of a set of initial conditions, its performance is highly dependent on the numerical integration method adopted into the procedure. Several integration methods were tested, like the Radau or Bulirsch–Stoer, and the best efficiency was obtained with a Runge-Kutta method of order 8 (DOP853) \citep[e.g.][]{bookrk8}.
The computational speed of the TFC and NM procedures were tested on a machine equipped with an i7-4790 processor. To analyze the speed of the methods, the computational time required to retrieve the 324 area-to-mass ratios from a set of simulated observational data (used to generate Fig. \ref{fig:diag2_re}) was measured. The code based in the TFC retrieved the 324 values of the area-to-mass parameter in 1.836634 seconds.
In contrast, the code based on the NM method retrieved the 324 values of the area-to-mass parameter values from the same set of simulated observational data (used for Fig. \ref{fig:diag2_re}) in 621.743514 seconds.

\section{Conclusions}
\label{sec:con}

This paper presents techniques for finding the parameters of the dynamical model of a satellite from observational data. The study adopts the TFC (Theory of Functional Connections) and NM (Nelder-Mead) simplex algorithm as optimization methods to determine the optimal area-to-mass ratio of a satellite using an inverse problem approach. The accuracy of these methods is evaluated for different altitudes and initial conditions, and the results are compared with those generated using GMAT (General Mission Analysis Tool).
The Nelder-Mead method is chosen as one of the optimization methods due to its suitability for solving the orbit determination problem with a moderate number of parameters and its ability to handle noisy objective functions, which is crucial when dealing with real-world observations and measurements. On the other hand, TFC is selected for its efficiency, potential in solving complex optimization problems with a large number of variables, and its ability to incorporate observational data into the dynamical model.


Comparisons with observational data simulated using our data generator demonstrate the highest level of accuracy achievable by these methods when the constraints align with the dynamical model. In this scenario, the TFC method exhibits higher accuracy than the NM method, with the retrieved parameter being orders of magnitude closer to the original parameter. However, this high level of accuracy may not be significant when the observational data does not perfectly fit the dynamical model, as is the case with real satellites subject to imperfectly known perturbations. To address this problem, the paper analyzes the case where the constraints are not perfect solutions of the dynamical model using observational data simulated with GMAT. In this case, both the TFC and NM methods demonstrate a similar and reasonably good accuracy, albeit inferior to the observational data generated by our data generator. This implies that both the TFC and NM methods proposed in this paper have the potential to accurately retrieve the parameters of a real satellite, primarily depending on external factors such as the quality of the observational data and the dynamical model, rather than the method itself.
The paper also examines the computational resource consumption of the methods. The TFC procedure has shown a few orders of magnitude faster convergence than the NM method to retrieve the parameter.

\section*{Acknowledgements}
We acknowledge support from the National Institute for Space Research - INPE. This work is supported by the European Regional Development Fund
(FEDER), through the Competitiveness and Internationalization
Operational Programme (COMPETE 2020) of the Portugal 2020 framework
[Project SmartGlow with Nr. 069733 (POCI-01-0247-FEDER-069733)]; The
team acknowledges further support from ENGAGE-SKA Research
Infrastructure, ref. POCI-01-0145-FEDER-022217, funded by COMPETE 2020
and FCT, Portugal; IT team members acknowledge support from Projecto
Lab. Associado UID/EEA/50008/2019.  We acknowledge  support by the
European Commission H2020 Programme under the grant agreement 2-3SST2018-20. Support from Center for Mechanical and Aerospace Science and Technologies - C-MAST, funded by FCT-Fundação para a Ciência e a Tecnologia through project UIDB/00151/2020. This work is also supported by CFisUC (UIDB/04564/2020 and UIDP/04564/2020). This work is supported by the Fundação para a Ciência e Tecnologia (FCT), Ph.D. grant No.2022.12341.BDANA.

\bibliography{mybib}

\appendix
\section{Evaluation of the acceleration $\textit{g}_{10}$ for the point $(x,y,z)$}
\label{app:A}

The acceleration $\B{g}_{N}$ up to the order $N$ for the point of coordinates $(x,y,z)$ in the ITRF is given by
\begin{equation}
\B{g}_{N}=\vb{\nabla}V=\fracpar{V}{r}\vb{u}_r+\frac{1}{r}\fracpar{V}{\t}\vb{u}_\t+\frac{1}{r\sin \t}\fracpar{V}{\lambda}\vb{u}_\lambda,
\end{equation}
with the spherical coordinates $(r,\t,\lambda)$ where $\t$ is the colatitude, the orthonormal spherical basis $(\vb{u}_r,\vb{u}_\t,\vb{u}_\lambda)$, and the potential $V$ given by
\begin{equation}
V\left(r,\t,\lambda\right)=\frac{GM}{r}\left(1+\sum^N_{n=2}\left(\frac{a}{r}\right)^{n}\sum^n_{m=0}\left(\C\cos m\lambda+\S\sin m\lambda\right)\P\left(\ct\right)\right),
\end{equation}
for the order $n$, the degree $m$, the normalized spherical harmonic coefficients of the gravitational potential $(\C,\S)$, the normalized associated Legendre polynomials $\P$, and $\ct=\cos\theta$

By definition, the Legendre polynomials $P_n$ are given by \citep[e.g.][]{1972hmfw.book.....A}
\begin{equation}
P_n\left(\ct\right)=\frac{1}{2^n}\sum^{\left\lfloor\frac{n}{2}\right\rfloor}_ {l=0}\left(-1\right)^l\left(\begin{matrix} n \\ l \end{matrix}\right)\left(\begin{matrix} 2n-2l \\ n \end{matrix}\right)\ct^{n-2l},
\end{equation}
where $\lfloor\rfloor$ is the floor function, and the normalized associated Legendre polynomials $\P$ by
\begin{equation}
\P\left(\ct\right)=\frac{1}{\N}\left(1-\ct^2\right)^{\frac{m}{2}}\frac{d^m}{d\ct^m}P_n\left(\ct\right),
\end{equation}
with
\begin{equation}
\N=\sqrt{\frac{\left(n+m\right)!}{\left(n-m\right)!\left(2n+1\right)\left(2-\delta_{0m}\right)}},
\end{equation}
and the Kronecher delta $\delta$.
The normalized associated Legendre polynomials $\P$ can be then expressed as
\begin{equation}
\P\left(\ct\right)=\left(1-\ct^2\right)^{\frac{m}{2}}\sum^{\left\lfloor\frac{n-m}{2}\right\rfloor}_ {j=0}\gamma_{nmf\left(j\right)}\ct^{f\left(j\right)},
\end{equation}
where
\begin{equation}
\gamma_{nml}=\frac{\prod^{n}_{p=1}\left(n+m+l+1-2p\right)}{l!\left(n-l-m\right)!\N},
\end{equation}
and
\begin{equation}
f\left(j\right)=2j+\left\lfloor\frac{n+m+1}{2}\right\rfloor-\left\lfloor\frac{n+m}{2}\right\rfloor.
\end{equation}

The derivatives of the potential with the respect to the spherical coordinates are then given with $\r=r/a$ by
\begin{equation}
\fracpar{V}{r}=  -\frac{GM}{a^2\r^2}\Bigg(1+\sum^N_{n=2}\frac{1}{\r^n}\sum^n_{m=0}\left(1-\ct^2\right)^{\frac{m}{2}}\sum^{\left\lfloor\frac{n-m}{2}\right\rfloor}_{j=0}\left(\a_{1nmf\left(j\right)}\cos m\lambda+\b_{1nmf\left(j\right)}\sin m\lambda\right)\ct^{f\left(j\right)}\Bigg),
\end{equation}
\begin{equation}
\fracpar{V}{\t}= \frac{GM}{a\r}\Bigg(\sum^N_{n=2}\frac{1}{\r^n}\sum^n_{m=0}\left(1-\ct^2\right)^{\frac{m-1}{2}}\sum^{\left\lfloor\frac{n+1-m}{2}\right\rfloor}_{j=0}\left(\a_{2nmg\left(j\right)}\cos m\lambda +\b_{2nmg\left(j\right)}\sin m\lambda\right)\ct^{g\left(j\right)}\Bigg),
\end{equation}
\begin{equation}
\fracpar{V}{\lambda}= \frac{GM}{a\r}\Bigg(\sum^N_{n=2}\frac{1}{\r^n}\sum^n_{m=1}\left(1-\ct^2\right)^{\frac{m}{2}}\sum^{\left\lfloor\frac{n-m}{2}\right\rfloor}_{j=0}\left(\a_{3nmf\left(j\right)}\cos m\lambda +\b_{3nmf\left(j\right)}\sin m\lambda\right)\ct^{f\left(j\right)}\Bigg),
\end{equation}
with
\begin{equation}
g\left(j\right)=2j+\left\lfloor\frac{n+m}{2}\right\rfloor-\left\lfloor\frac{n+m-1}{2}\right\rfloor,
\end{equation}
and the coefficients
\begin{equation}
\a_{1nml}=\gamma_{nml}\left(n+1\right)\C, \ \qquad \b_{1nml}=\gamma_{nml}\left(n+1\right)\S,
\end{equation}
\begin{equation}
\a_{2nml}=\epsilon_{nml}\C, \ \qquad \b_{2nml}=\epsilon_{nml}\S,
\end{equation}
\begin{equation}
\a_{3nml}=\gamma_{nml}m\S, \ \qquad \b_{3nml}=-\gamma_{nml}m\C,
\end{equation}
with
\begin{equation}
\epsilon_{nml}=\frac{\left(n+1-m\right)\left(n+m+l\right)-l\left(n+1\right)}{l!\left(n+1-l-m\right)!}\left(\prod^{n}_{p=1}\left(n+m+l-2p\right)\right).
\end{equation}

\end{document}